\documentstyle[preprint,pra,aps]{revtex}

\begin{document}

\bibliographystyle{prsty}
\draft
\tighten
 
\title{Fidelity and information in the quantum teleportation of 
continuous variables}

\author{Holger F. Hofmann, Toshiki Ide, and Takayoshi Kobayashi}
\address{Department of Physics, Faculty of Science, University of Tokyo,\\
7-3-1 Hongo, Bunkyo-ku, Tokyo113-0033, Japan}

\author{Akira Furusawa}
\address{Nikon Corporation, R\&D Headquarters,\\
Nishi-Ohi, Shinagawa-ku, Tokyo 140-8601, Japan}

%\author{Holger F. Hofmann\\Department of Physics, Faculty of Science, 
%University of Tokyo\\7-3-1 Hongo, Bunkyo-ku, Tokyo113-0033, Japan}

\date{\today}

\maketitle

\begin{abstract}
Ideally, quantum teleportation should transfer a quantum state without
distortion and without providing any information about that state.
However, quantum teleportation of continuous electromagnetic field 
variables introduces additional noise, limiting the
fidelity of the quantum state transfer. In this article, the 
operator describing the quantum state transfer is derived. The transfer
operator modifies the probability amplitudes of the quantum state in a
shifted photon number basis by enhancing low photon numbers and suppressing
high photon numbers. This modification of the statistical weight corresponds
to a measurement of finite resolution performed on the original quantum state.
The limited fidelity of quantum teleportation is thus shown to be a direct
consequence of the information obtained in the measurement. 
\end{abstract}

\section{introduction}
Quantum teleportation is a process by which the quantum state of a system
A can be transfered to a remote system B by exploiting the entanglement
between system B and a reference system R. Ideally, no information is obtained
about system A, even though the exact relationship between A and R is 
determined by measuring a set of joint properties of A and R. While the
original state of A is lost in this measurement, it can be recovered by 
deducing the relationship between A and B from the original entanglement
between B and R and the measured entanglement between A and B. 

\begin{figure}
\begin{picture}(400,300)
%\put(0,0){\framebox(400,300){}}

\put(160,40){\framebox(80,40){\Large OPA}}

\put(160,85){\line(-1,1){45}}
\put(155,80){\line(-1,1){45}}
\put(110,130){\line(0,-1){10}}
\put(110,130){\line(1,0){10}}
\put(135,105){\makebox(20,20){\Large R}}

\put(240,85){\line(1,1){45}}
\put(245,80){\line(1,1){45}}
\put(290,130){\line(0,-1){10}}
\put(290,130){\line(-1,0){10}}
\put(245,105){\makebox(20,20){\Large B}}

\put(40,85){\line(1,1){45}}
\put(45,80){\line(1,1){45}}
\put(90,130){\line(0,-1){10}}
\put(90,130){\line(-1,0){10}}
\put(45,105){\makebox(20,20){\Large A}}
\put(10,66){\makebox(40,12){\large Input}}
\put(10,54){\makebox(40,12){\large field}}

\put(100,110){\line(0,1){60}}
\put(80,90){\makebox(40,12){\large Beam}} 
\put(80,78){\makebox(40,12){\large splitter}}

\put(90,155){\line(-1,1){45}}
\put(85,150){\line(-1,1){45}}
\put(40,200){\line(0,-1){10}}
\put(40,200){\line(1,0){10}}
\put(30,200){\makebox(20,20){\Large $x_-$}}

\put(110,155){\line(1,1){45}}
\put(115,150){\line(1,1){45}}
\put(160,200){\line(0,-1){10}}
\put(160,200){\line(-1,0){10}}
\put(150,200){\makebox(20,20){\Large $y_+$}}

\put(40,230){\framebox(120,60){}}
\put(60,260){\makebox(80,20){\large Measurement of}}
\put(60,240){\makebox(80,20){\large $\beta=x_-+i y_+$}}

\bezier{400}(160,260)(250,260)(290,180)
\put(290,180){\line(0,1){12}}
\put(290,180){\line(-3,2){10}}

\put(295,135){\framebox(40,40){\Large $\hat{D}(\beta)$}}

\put(340,185){\line(1,1){25}}
\put(345,180){\line(1,1){25}}
\put(370,210){\line(0,-1){10}}
\put(370,210){\line(-1,0){10}}

\put(350,227){\makebox(40,12){\large Output}}
\put(350,215){\makebox(40,12){\large field}}

\end{picture}
\caption{\label{setup} Schematic representation of the quantum teleportation 
scheme.}
\end{figure}
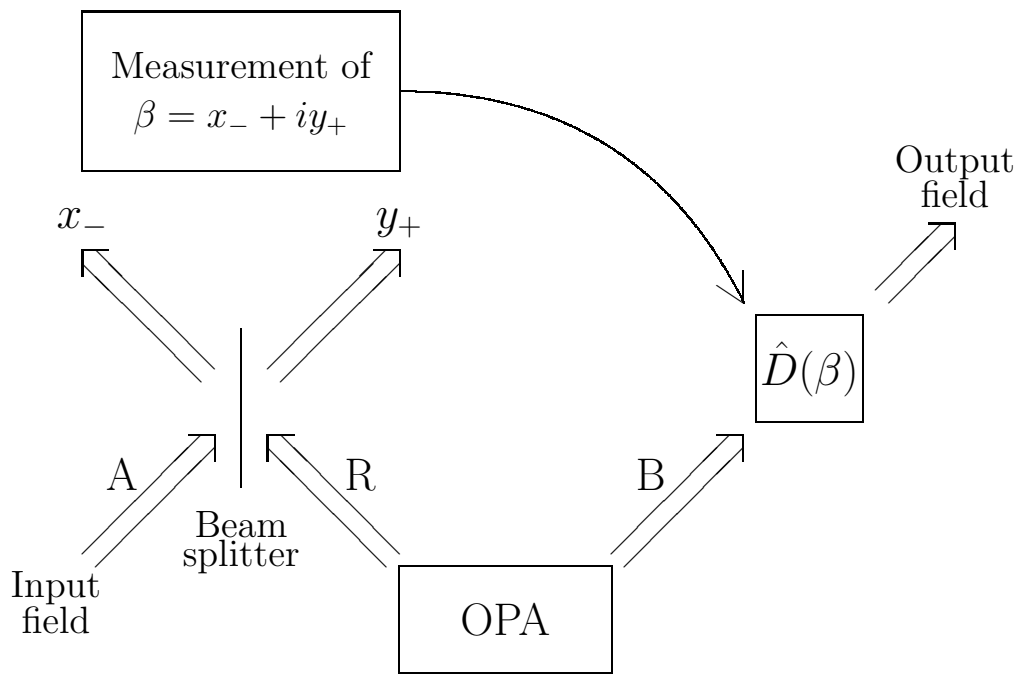

The original proposal of quantum teleportation \cite{Ben93} assumed maximal 
entanglement between B and R. However, it is also possible to realize
quantum teleportation with non-maximal entanglement. In particular, such a
teleportation scheme has been applied to the quantum states of light field
modes \cite{Bra98,Fur98}, 
where maximal entanglement is impossible since it would 
require infinite energy. A schematic setup of this scheme is shown in figure 
\ref{setup}. This approach to quantum teleportation has inspired a number
of investigations into the dependence of the teleportation process on the
details of the physical setup \cite{Mil99,Opa00,Hor00}. 
In this context, it is desirable to develop compact theoretical 
formulations describing the effects of this teleportation scheme on the 
transferred quantum state.

Originally, the teleportation process for continuous variables has been
formulated in terms of Wigner functions \cite{Bra98}. Recently, a 
description in the discrete photon number base has been provided as
well \cite{Enk99}. In the following, the latter approach will be reformulated
using the concept of displaced photon number states, and a general transfer 
operator $\hat{T}(x_-,y_+)$ will be derived for the quantum teleportation of a 
state associated with a measurement result of $x_-$ and $y_+$. 
This transfer operator describes the modifications which the quantum state
suffers in the teleportation as well as the information obtained about the 
quantum state due to the finite entanglement. It is shown that this type
of quantum teleportation resembles a non-destructive measurement of light 
field coherence with a measurement resolution given by the entanglement
of B and R.

\section{Measuring the entanglement of unrelated field modes}
The initial step in quantum teleportation requires a measurement
of the entanglement between input system A and reference system R.
Ideally, this projective measurement does not provide any information 
about properties of A by itself. 

In the case of continuous field variables \cite{Bra98,Fur98}, the measured
variables are the difference $\hat{x}_-=\hat{x}_A-\hat{x}_R$ and the sum 
$\hat{y}_+=\hat{y}_A+\hat{y}_R$ of the orthogonal quadrature components. The 
eigenstates of these two commuting variables may be expressed in terms of 
the photon number states $\mid n_A;n_R \rangle $ as
\begin{eqnarray}
\label{eq:rep}
\mid \beta (A,R) \rangle &=& \frac{1}{\sqrt{\pi}}
\sum_{n=0}^{\infty} \hat{D}_A(\beta) \mid n ; n \rangle
\nonumber \\
\mbox{with} && \hat{x}_- \mid \beta (A,R) \rangle 
= \mbox{Re}(\beta) \mid \beta (A,R) \rangle
\nonumber \\
\mbox{and} && \hat{y}_+ \mid \beta (A,R) \rangle 
= \mbox{Im}(\beta) \mid \beta (A,R) \rangle
,
\end{eqnarray}
where the operator $\hat{D}_A(\beta)$ is the displacement operator 
acting on the input field A, such that 
\begin{eqnarray}
\hat{D}_A(\beta) &=& 
\exp\left(2i \mbox{Im}(\beta) \hat{x}_A-2i \mbox{Re}(\beta) \hat{y}_A\right)
\nonumber \\ 
\mbox{with} &&
\hat{D}^\dagger_A(\beta) \hat{x}_A \hat{D}_A(\beta) 
= \hat{x}_A + \mbox{Re}(\beta)  
\nonumber \\ 
\mbox{and} &&
\hat{D}^\dagger_A(\beta) \hat{y}_A \hat{D}_A(\beta) 
= \hat{y}_A + \mbox{Im}(\beta)
.
\end{eqnarray}
Of course, the coherent shift could also be applied to field R instead
of field A. However, in the representation given by equation (\ref{eq:rep}),
it is easy to identify the quantum state associated with a photon number
of the reference field R. 

If the quantum state $\mid \psi_R \rangle$ 
of the reference field R is known, the measurement 
result provides information on the quantum state $\mid \psi_A \rangle$ 
of field A through the probability distribution $P(\beta)$ given by
\begin{eqnarray}
P(\beta) &=& \frac{1}{\pi} |
\sum_{n=0}^{\infty} \langle \psi_A\mid \hat{D}_A(\beta)\mid n \rangle
                    \langle \psi_R\mid n \rangle
|^2 
\nonumber \\
&=& \frac{1}{\pi} | \langle \psi_A\mid \hat{D}_A(\beta)\mid \psi_R^* \rangle
|^2,
\nonumber \\
\mbox{where} &&
\mid \psi_R^*\rangle = \sum_{n=0}^{\infty} \langle n \mid \psi_R \rangle^* 
\mid n \rangle.
\end{eqnarray}
Effectively, the measurement of entanglement projects the quantum state of
field A onto a complete non-orthogonal measurement basis given by the displaced
reference states $\mid \psi_R^*\rangle$. The completeness of this measurement
basis is given by
\begin{equation}
\frac{1}{\pi}\int d^2\!\beta \; \hat{D}(\beta) \mid \psi_R^*\rangle
                          \langle \psi_R^*\mid \hat{D}^\dagger(\beta) 
= \hat{1}.
\end{equation}
In the case of ``classical'' teleportation, the reference field R is in the
quantum mechanical vacuum state $\mid n=0 \rangle$. Therefore, the measurement
of the field entanglement given by $\beta$ projects the incoming signal
field A directly onto a displaced vacuum state. 

\section{Quantum teleportation} 
In the general case of quantum teleportation, the quantum state of the
reference field R cannot be determined locally because of its entanglement 
with the remote field B. This indicates that the type of measurement 
performed is unknown until the remote system is measured as well. The 
meaning of the measurement result $\beta$ depends on the unknown
properties of the remote field B. The entangled reference field R thus 
provides the means to choose between complementary measurement types even
after the measurement interaction between input field A and reference
field R has occurred. 

Within the quantum state formalism, the initial state of the entangled
fields R and B may be written as \cite{Enk99,Wal94}
\begin{equation}
\mid q(R,B) \rangle = \sqrt{1-q^2} \sum_{n=0}^{\infty} q^{n} \mid n;n \rangle.
\end{equation}
Thus, the photon numbers of the reference field R and the remote field B
are always equal. However, low photon numbers are more likely than high 
photon numbers, so the two mode entanglement is limited by the information 
available about the photon number of each mode. In a measurement of the
entanglement between field A and field R, this information about R is 
converted into measurement information about A, thus causing a decrease
in fidelity as required by the uncertainty principle.

A measurement of the entanglement between field A and field R projects
the product state $\mid \psi_A \rangle \otimes \mid q(R,B)\rangle$ into
a quantum state of the remote field B given by
\begin{equation}
\mid \psi_B (\beta)\rangle = \sqrt{\frac{1-q^2}{\pi}} \sum_{n=0}^{\infty}
q^{n} \mid n\rangle \langle n \mid \hat{D}_A(-\beta) \mid \psi_A \rangle,
\end{equation}
where the measurement probability $P(\beta)$ is given by 
$\langle \psi_B(\beta) \mid \psi_B (\beta)\rangle$. Thus the measurement 
determines the displacement $\beta$ between field A and field B, resulting
in a quantum state $\mid \psi_B (\beta)\rangle$ that appears to be a
copy of the input state $\mid \psi_A \rangle$, displaced by $-\beta$.
However, the measurement information obtained because low photon numbers
are more likely than high photon numbers in both R and B causes a statistical
modification of the probability amplitudes of the photon number states in
the remote field B. 

The final step in quantum teleportation is the reconstruction of the initial
state from the remote field by reversal of the displacement. The output state
then reads
\begin{eqnarray}
\mid \psi_{\mbox{out}} (\beta)\rangle &=&  \hat{T}(\beta) \mid \psi_A \rangle
\nonumber \\
\mbox{with} &&
\hat{T}(\beta) = \sqrt{\frac{1-q^2}{\pi}} \sum_{n=0}^{\infty}
q^{n} \hat{D}_A(\beta) \mid n\rangle \langle n \mid \hat{D}_A(-\beta).
\end{eqnarray}
Note that this output state is not normalized because 
$\langle\psi_{\mbox{out}} (\beta)\mid \psi_{\mbox{out}} (\beta)\rangle$
defines the probability of the measurement results $\beta$.
The complete process of quantum teleportation is thus summed up by the 
transfer operators $\hat{T}(\beta)$. 	

\section{Transfer operator properties}
The transfer operator $\hat{T}(\beta)$ determines not only the properties 
of the quantum state after the teleportation process following a measurement
result of $\beta$ for the field entanglement between A and R, but also the 
probability of obtaining the result $\beta$ itself. The probability 
distribution $P(\beta)$ is given by
\begin{eqnarray}
P(\beta) &=& \langle \psi_A \mid \hat{T}^2(\beta) \mid \psi_A \rangle
\nonumber \\
         &=& \frac{1-q^2}{\pi}\sum_{n=0}^{\infty} q^{2n} 
                 |\langle n \mid \hat{D}_A(-\beta) \mid \psi_A \rangle|^2.
\end{eqnarray}
Since the prefactor $q^n$ is larger for small n, a measurement result of 
$\beta$ is more likely if the photon number of the displaced state
$\hat{D}_A(-\beta) \mid \psi_A \rangle$ is low. It is possible to identify
the displaced photon number with the square of the field difference between
$\beta$ and the actual field value of A. Therefore, a measurement
result of $\beta$ makes large deviations of the field A from this value
of $\beta$ unlikely.

The transfer operator $\hat{T}(\beta)$ also determines the relationship
between the input state and the output state. Since it is the goal of
quantum teleportation is to achieve identity between the input state
and the output state, the overlap between the two states may be used
as a measure of the fidelity of quantum teleportation. For a single
teleportation event associated with a measurement result of $\beta$,
this fidelity is given by
\begin{equation}
\label{eq:fid}
F(\beta) = \frac{1}{P(\beta)}
|\langle \psi_A \mid \hat{T}(\beta) \mid \psi_A \rangle|^2.
\end{equation}
For instance, the fidelity of quantum teleportation for a photon
number state displaced by $\beta$ is exactly 1. However, it is unlikely
that $\beta$ will be exactly equal to the displacement of the photon number
state to be teleported, so the average fidelity will be much lower.
The average fidelity $F_{\mbox{av.}}$ is given by
\begin{eqnarray}
F_{\mbox{av.}} &=& \int d^2\!\beta\; P(\beta) F(\beta)
\nonumber \\
               &=& \int d^2\!\beta\; 
|\langle \psi_A \mid \hat{T}(\beta) \mid \psi_A \rangle|^2.
\end{eqnarray}
Since the transfer operator $\hat{T}(\beta)$ is different for each 
teleportation event, the output field states show unpredictable fluctuations.
These fluctuations may be expressed in terms of a density matrix,
\begin{equation}
\hat{\rho}_{\mbox{out}}= \int d^2\!\beta\; \hat{T}(\beta)\mid \psi_A \rangle
                                        \langle \psi_A \mid \hat{T}(\beta).
\end{equation}
In terms of this mixed state density matrix, the average fidelity reads
\begin{equation}
F_{\mbox{av.}} = \langle \psi_A \mid \hat{\rho}_{\mbox{out}}
\mid \psi_A \rangle.
\end{equation}
However, the measurement information $\beta$ is available as classical 
information, so the density matrix $\hat{\rho}_{\mbox{out}}$ actually
underestimates the information available about the output field.
In particular, a verifier checking the fidelity of the transfer in B
can know the exact output state based on the knowledge of the input state
and the measurement result $\beta$.

\section{Fidelity and information}
The transfer operator $\hat{T}(\beta)$ describes how
the information $\beta$ obtained about the properties of the input state 
makes contributions from displaced photon number states less
likely as the displaced photon number increases. 
The changes in the quantum state which are responsible for a
fidelity below one correspond to this change in the statistical weight
of the quantum state components. This situation is typical for 
quantum mechanical measurements. It is impossible to obtain
information beyond the uncertainty limit without introducing a 
corresponding amount of noise into the system, because the probability
amplitudes correspond to both statistical information and to physical fact
\cite{Hof00}.
Simply by making one quantum state component more likely than another, 
the coherence between the two quantum state components is diminished and 
thus the noise in any variable depending on this coherence is increased.

In order to clarify the measurement information obtained, it is convenient
to represent the transfer operator $\hat{T}(\beta)$ in terms of coherent 
states. This representation is easy to obtain by using the formal analogy
of $\hat{T}(\beta)$ with a thermal photon number distribution. The result 
reads
\begin{equation}
\hat{T}(\beta) = \sqrt{\frac{1-q^2}{\pi^3 q^2}} 
\int d^2\!\alpha 
\exp\left(-\left(\frac{1-q}{q}\right) |\alpha-\beta|^2 \right)
\mid \alpha \rangle \langle \alpha \mid.
\end{equation}  
In the limit of $q\to 0$, the transfer operator thus corresponds to 
a projection operator on the coherent state $\mid \beta \rangle$. As 
$q$ increases, the operator corresponds to a mixture of weighted 
projections which prefer coherent states with field values close to 
$\beta$, distorting the field distribution of the input state. 

This result highlights the epistemological nature of the quantum state.
If one were to attribute physical reality to the quantum state,
the change of the quantum state given by $\hat{T}(\beta)$ would be 
a real physical effect and the statistical nature of the measurement 
result $\beta$ appears to be a rather arbitrary limitation. Why is it 
then not possible to control the force which molds the remote wave function?
If one assumes that only observable physical properties are real, however, 
then the quantum mechanical 
probabilities behave in close analogy to classical ones, permitting a 
much clearer understanding of both possibilities and limitations. The 
information gained about the input field $A$ due to the finite fluctuations 
in the reference field $R$ can then be considered as a 
valid measurement of the coherent field components. Since the resolution 
of the field information obtained is limited, the output state retains some 
of the quantum coherent properties of the original input such as squeezing or
cat state coherence. By performing measurements of physical properties other 
than the coherent field on the output state, one then obtains a mix of 
information relating back to the original input field.

In the light of this interpretation, quantum teleportation represents a 
delayed choice measurement, where the final selection of measurement variables
is performed when the remote field $B$ is measured, thus indirectly 
determining the physical properties of the reference field $R$ before 
the measurement of
$\beta$. The ``teleportation'' effect is then a purely classical information
transfer made possible by the statistical correlations of the relevant 
physical properties. The quantum nature of the procedure only emerges when the
noise properties of the initial and final states are compared. In particular,
this viewpoint may help to illustrate the nature of the dividing line 
between ``quantum teleportation'' and ``classical teleportation''
\cite{Bra98,Fur98}.

\section{Verification of quantum state statistics}
The output of a single teleportation process involving a pure state input
$\mid \psi_A \rangle$ results in a well defined pure state output 
$\hat{T}(\beta)\mid \psi_A \rangle$. Since the statistical properties
of this state are modified by the information obtained in the measurement
of $\beta$, the output state is different from the input state, as given
by the fidelity $F(\beta)$ defined by equation (\ref{eq:fid}). However,
this difference shows only in the statistics obtained by measuring the
output of an ensemble of identical input states. In other words, there is
no single measurement to tell us whether the output quantum state is 
actually identical to the input state. Non-orthogonal quantum states may
always produce the same measurement results. The verification process 
following the quantum teleportation is therefore a non-trivial process
requiring the comparison of measurement statistics which are generally
noisy, even for a fidelity of one. 

In the experimentally realized
teleportation of continuous variables reported in \cite{Fur98}, the 
verification is achieved by measuring one quadrature component 
of the light field using homodyne detection and comparing the result with
the quadrature noise of the coherent state input. 
This type of verification 
can be generalized as a projective measurement on a set of states 
$\mid V \rangle$ satisfying the completeness condition for positive valued
operator measures,
\begin{equation}
\sum_V \mid V \rangle \langle V \mid = \hat{1}.
\end{equation}
The probability of obtaining a verification result $V$ is given by
\begin{equation}
P(V) = \int d^2\!\beta\; |\langle V \mid \hat{T}(\beta) \mid \psi_A \rangle|^2.
\end{equation}
This probability distribution is then compared with the input distribution 
of the verification variable $V$. However, the total process of teleportation
and verification may be summarized in a single measurement of $\beta$ and
$V$. If the information inherent in the measurement result $\beta$ is
retained, the complete measurement performed on the input state is defined
by the projective measurement basis $\mid \beta, V \rangle$ given by
\begin{eqnarray}
\mid \beta, V \rangle = \hat{T}(\beta)\mid V \rangle.
\end{eqnarray}
The probability distribution over measurement results $\beta$ and verification
results $V$ then reads
\begin{eqnarray}
P(\beta,V) &=& |\langle \beta, V \mid \psi_A\rangle|^2
\nonumber \\
&=& |\langle V \mid \hat{T}(\beta) \mid \psi_A\rangle|^2.
\end{eqnarray}
The quantum measurement effectively performed on the input state is 
thus composed of the measurement step of quantum teleportation and
the verification step. The fidelity is determined by the difference
in the statistics over $V$ between this two step measurement and a 
direct measurement of $V$ only. However, the information lost in quantum
teleportation is actually less than is suggested by the average Fidelity.
If the teleportation result $\beta$ is considered as well, the combination 
of teleportation and verification extract the maximal amount of measurement
information permitted in quantum mechanics and consequently allows a complete
statistical characterization of the original input state. 

A particularly striking example can be obtained for a verification scheme
using eight port homodyne detection. In this case, the verification variable
is the coherent field $\alpha$ and the verification states  
$\mid V \rangle = 1/\sqrt{\pi} \mid \alpha \rangle$ are the associated coherent
states. The effective measurement basis is then given by
\begin{eqnarray}
\mid \beta, \alpha \rangle &=& \frac{1}{\sqrt{\pi}}
\hat{T}(\beta) \mid \alpha \rangle
\nonumber \\
&=& \frac{\sqrt{1-q^2}}{\pi} \exp\left(-(1-q^2)\frac{|\alpha-\beta|^2}{2}
\right)  
\mid \gamma = \beta + q (\alpha-\beta) \rangle.
\end{eqnarray}
The measurement still projects on a well defined coherent state, but the
coherent field $\gamma$ is a function of both the teleportation measurement
$\beta$ and the verification measurement $\alpha$.
It is therefore possible to reconstruct the correct measurement statistics 
of the input state by referring to the teleportation results $\beta$ as well
as to the verification results $\alpha$.  

\section{Conclusions}
In conclusion, the quantum teleportation of an input 
state $\mid \psi_A \rangle$
can be described by a measurement dependent transfer operator $\hat{T}(\beta)$
which modifies the quantum state statistics according to the information
obtained about the input state $\mid \psi_A \rangle$ in the measurement of 
$\beta$. The statistics of subsequent verification measurements may be derived
by directly applying the transfer operator $\hat{T}(\beta)$ to the states
$\mid V \rangle$ describing the projective verification measurement. 
Quantum information is only lost because the effective measurement basis 
$\mid \beta, V \rangle$ does not usually correspond to the eigenstate basis 
in which the information has been encoded. 

While the type of physical information which can be obtained about the 
original input field is restricted because some information necessarily 
``leaks out'' in quantum teleportation with limited entanglement, 
the total information obtained after the verification 
step still corresponds to the information obtained in an ideal projective 
measurement. The limitations imposed on quantum teleportation by a fidelity
less than one are thus a direct consequence of the measurement information
obtained about the transfered state in the measurement of $\beta$ and can
be applied directly to the measurements performed after the transfer of the
quantum state.

\section*{Acknowledgements}
One of us (HFH) would like to acknowledge support from the Japanese 
Society for the Promotion of Science, JSPS.

%=========================================================

%=========================================================

\end{document}